\documentclass[useAMS,usenatbib]{rmaa}

\begin{document}

\title{Functional relationships for $T_{\rm \lowercase {eff}}$ and \lowercase {log}~$\lowercase {g}$ in F-G supergiants from $\lowercase {uvby}\beta$ photometry}

\author{A. Arellano Ferro\altaffilmark{1}
\medskip }

\altaffiltext{1}{Instituto de
Astronom\'{\i}a, Universidad Nacional Aut\'onoma de M\'exico.}

\fulladdresses{
\item A. Arellano Ferro: Instituto de Astronom\'{\i}a,
UNAM, Apdo. Postal 70-264, M\'exico, D.F. M\'exico
(armando@astro.unam.mx).}

\shortauthor{A. Arellano Ferro}
\shorttitle{Temperature and gravity from $\lowercase{uvby}\beta$ in F-G supergiants}

\resumen{A partir de datos fotoel\'ectricos en el sistema $uvby\beta$ y de valores precisos sint\'eticos y espectrosc\'opicos de $T_{\rm eff}$ y log~$g$ en 50 supergigantes de tipos espectrales F-G, hemos calculado relaciones funcionales que permiten la estimaci\'on inicial
de la temperatura efectiva y la gravedad en este tipo de estrellas. Se demuestra que aunque las calibraciones
de $T_{\rm eff}$ fueron calculadas con los datos de estrellas
supergigantes j\'ovenes y masivas, tambi\'en son v\'alidas para estrellas evolucionadas post-AGB y RV Tau de temperaturas similares. La gravedad superficial tambi\'en puede calcularse a partir del \1ndice
$\Delta [c_1]$ con una precisi\'on de 0.26 dex. Aunque se puede distinguir una correlaci\'on entre 
$M_V$ y $\Delta [c_1]$, no hemos encontrado una correlaci\'on que prediga $M_V$ de manera suficientemente precisa.}

\abstract{From photoelectric $uvby\beta$ data and recent accurate synthetic and spectroscopic values of $T_{\rm eff}$ and log~$g$ for 50 F-G supergiants, we have calculated functional relationships that
lead to initial estimates of effective temperature and gravities for these types of stars. It is shown that 
while the $T_{\rm eff}$ relationships are calculated using the data on young massive supergiants, they are
also valid for evolved stars of similar temperatures like post-AGB and RV Tau stars. The gravity can also be predicted from the $\Delta [c_1]$ index with an uncertainty of about 0.26 dex. Although a clear and significant trend between $M_V$ and $\Delta [c_1]$ is seen, no calibration is found that predicts accurate values of $M_V$.}


\maketitle

\section{INTRODUCTION}

Back in the mid 1990's we had calculated a set of functional relationships between several reddening-free indices in the Str\"omgren $uvby\beta$ system and the effective temperature
for 41 supergiants of luminosity classes I and II and with spectral types between A0 and K0. The main goal of those relationships was to provide initial values of the effective temperature for our own spectroscopic work aimed to calculate
detailed atmospheric abundances for stars of intermediate temperatures. At that time we used
the temperatures calculated by Bravo-Alfaro et al. (1997) from 13-color photometry and the 
$uvby\beta$ data from the catalogue of Arellano Ferro et al. (1998).

The relationships turned out to be quite useful for several problems related to the estimation of the temperature for these types of stars. However they remained unpublished mainly because the intrinsic scatter was 
rather large, probably due to the limited quality of the temperature data used.

Very recently a set of accurate temperatures, gravities and distances for 48 near (d $\leq$ 700 pc) and 15 distant (d $\geq$ 700 pc) A-G supergiants, were published (Lyubimkov et al. 2010). In the present paper functional relationships are worked out
in the light of these new values of physical parameters with the aim to provide a tool to
estimate initial values of $T_{\rm eff}$ and log~$g$ in yellow supergiants for subsequent spectroscopic work.

The paper is organized as follows: in  $\S$ \ref{data} the sources of the 
$uvby\beta$ data are described, in $\S$ \ref{temp} the calibrations for the temperature as 
a function of the [$c_1$], [$m_1$] and H$\beta$ photometric indices are discussed, in $\S$ \ref{grav} the gravity calibration in terms of the $\Delta [c_1]$ index is calculated,
in $\S$ \ref{pagb} we discuss the validity of the calibrations for post-AGB and RV Tau stars,
in $\S$ \ref{MV} the attempts of a calibration of the absolute magnitude and its limitations
are presented and in 
$\S$ \ref{concl} we summarize our conclusions.

\section {The photometric $\lowercase {uvby}\beta$ data}
\label{data}

The $uvby\beta$ photometric data have been taken from the catalogue of Arellano Ferro et al. (1998). For about a dozen of stars photometric data are not available in that source and their data have been obtained from the sources given in the Simbad database. When several measurements exist a simple average was calculated. In Table \ref{T1} we present for the sample stars thier reddening free indices 
[$c_1$] = $c_1$ -0.16 $(b-y)$ and [$m_1$] = $m_1$ + 0.33 $(b-y)$, and
their values of distance, effective temperature 
and gravity taken from Lyubimkov et al. (2010).

\begin{table*}
\footnotesize{
\begin{center}
\caption[{\small }] {\small Physical parameters and photometric indices for the sample F-G supergiants.}
\label{T1}
\hspace{0.01cm}
 \begin{tabular}{lrcccccccrcl}
\hline
 Star & D & $T_{\rm eff}$ & log~$g$ &[$c_1$]& [$m_1$]&$H\beta$&$E(b-y)$ & $M_V$&$\Delta [c_1]$ &Sp.T. & Comment \\
& (pc) & (K) & (dex)  &  & & & & & & \\
\hline
 HR27   &380. &6270.  &2.10   &1.022 & 0.222 &2.668&0.100&$-3.3$&  -0.236 &F2 II &    \\
 HR~157  &259. &5130.  &2.15   &0.293 & 0.504 &2.585&$-$0.031&$-1.5$ &-0.130 &G2.5 IIa & \\
 HR~207  &935. &5220.  &1.55   &0.380 & 0.145 &2.626&0.553 &$-3.9$ &0.000 &G0 Ib& \\ 
 HR~792  &397. &5020.  &2.09   &0.245 & 0.645 & &0.034&$-1.8$ & -0.062 &G5 II &D   \\     
 HR~849  &538. &5020.  &1.73   &0.209 & 0.707 &2.605&0.019&$-8.6$ & -0.068 &G5 Iab:&D   \\      
 HR~1017 &156. &6350.  &1.90   &1.048 & 0.300 &2.681&0.097&$-5.1$ &  0.114 &F5 Ib &V,S   \\  
 HR~1135 &170. &6560.  &2.44   &0.950 & 0.271 &2.684&0.046&$-2.6$ &  -0.093 &F5 II &V,RRL \\
 HR~1242 &629. &6815.  &1.87   &1.464 & 0.229 &2.730&0.310& $-5.3$&  0.236 &F0 II &   \\      
 HR~1270 &427. &5060.  &1.91   &0.260 & 0.635 &2.580&0.045&$-2.1$ &  -0.050 &G8 IIa &    \\ 
 HR~1303 &275. &5380.  &1.73   &0.451 & 0.479 &2.609&0.097&$-3.5$ &  -0.018 &G0 Ib &D,SB \\   
 HR~1327 & 98. &5440.  &2.89   &0.326 & 0.457 &2.574&$-$0.024&$+0.4$ &  -0.183 &G5 IIb &S \\      
 HR~1603 &265. &5300.  &1.79   &0.414 & 0.500 &2.607&0.002&$-0.6$  & -0.009 &G1 Ib-IIa &D \\   
 HR~1829 & 49. &5450.  &2.60   &0.366 & 0.423 &2.576&0.007&$ -0.6$&  -0.232 &G5 II &S   \\      
 HR~1865 &680. &6850.  &1.34   &1.451 & 0.202 &2.730&0.121& $-7.1$ & 0.086 &F0 Ib &V,S    \\ 
 HR~2000 &224. &5000.  &2.45   &0.304 & 0.519 &2.568&$-$0.045&$-0.4$ &  -0.105 &G2 Ib-II &S\\    
 HR~2453 &633. &4900.  &1.70   &0.224 & 0.648 &2.582&$-$0.033&$-2.4$ &  -0.083 &G5 Ib &D \\  
 HR~2597 &935. &6710.  &2.00   &1.333 & 1.777 &2.721&0.193 &$-4.3$ &-0.071 &F2 Ib-II & \\   
 HR~2693 &495. &5850.  &1.00   &0.869 & 0.446 &2.660&0.013&$-6.7$ &   0.337 &F8 Ia &V,S    \\ 
 HR~2786 &386. &5260.  &1.90   &0.367 & 0.556 &2.619&0.003&$-2.7$ &  -0.003 &G2 Ib &D,S   \\    
 HR~2833 &375. &5380.  &2.21   &0.391 & 0.535 &2.586&$-$0.010&$-1.8$ &  -0.005 &G3 Ib &    \\  
 HR~2881 &461. &5300.  &1.66   &0.413 & 0.557 &2.622&$-$0.029&$-3.6$ &  0.042 &G3 Ib &S    \\     
 HR~3045 &370. &4880.  &1.21   &0.251 & 0.793 &2.625&$-$0.030&$-4.4$ &  0.000 &G6 Ia &D,S  \\   
 HR~3073 &338. &6670.  &2.61   &1.145 & 0.271 &2.700&0.080&$-2.2$ &   0.106 &F1 Ia &   \\   
 HR~3102 &161. &5690.  &2.17   &0.520 & 0.395 &2.651&$-$0.005&$-1.8$ &  -0.144 &F7 II &S \\   
 HR~3188 &325. &5210.  &1.75   &0.325 & 0.569 &2.589&$-$0.026&$-3.1$ &  -0.035 &G2 Ib &D \\      
 HR~3229 &267. &5130.  &2.04   &0.281 & 0.663 &2.584&$-$0.003&$-2.1$ &  -0.016 &G5 II &  \\  
 HR~3291 &900. &6600.  &1.25   &1.413 & 1.996 &2.703& 0.224&$-7.2$ & 0.165& F3 Ib&D \\ 
 HR~3459 &236. &5370.  &2.08   &0.424 & 0.464 &2.609&0.005&$-2.3$ &  -0.072 &G1 Ib &D  \\     
 HR~4166 &177. &5475.  &2.36   &0.380 & 0.459 &2.586&$-$0.028&$-1.4$ &  -0.129 &G2.5 IIa &\\     
 HR~5143 &166. &5190.  &2.75   &0.277 & 0.540 &2.543&$-$0.023&$+0.2$ &  -0.113 &G5 II: & \\      
 HR~5165 &253. &5430.  &2.37   &0.480 & 0.416 &2.591&0.069&$-1.7$  & -0.135 &G0 Ib-IIa &   \\
 HR~6144 &330. &7400.  &1.80   &1.509 & 2.277 & &0.169&$-5.3$ &0.000 & A7 Ib &pAGB?\\  
 HR~6536 &117. &5160.  &1.86   &0.323 & 0.524 &2.598&0.016&$-2.6$ &  -0.359 &G2 Ib-IIa &D \\   
 HR~6978 &649. &6000.  &1.70   &0.780 & 0.383 &2.636&0.020&$-4.4$ &   0.075 &F7 Ib &    \\     
 HR~7014 &1000.&6760.  &1.66   &1.429 & 2.043 & &0.226&$-5.8$ & 0.195& F2 Ib& \\  
 HR~7094 &855. &6730.  &1.75   &1.208 &1.460  &2.711&0.227&$-5.2$ & -0.026&  F2 Ib& \\   
 HR~7264 &156. &6590.  &2.21   &1.025 & 0.242 &2.701&0.042&$-3.2$ &  -0.140 &F2 II &D,S \\   
 HR~7387 &880. &6700.  &1.43   &1.406 &1.976  &2.715&0.189&$-6.7$ &0.015 & F3 Ib& D  \\      
 HR~7456 &370. &5550.  &2.06   &0.459 & 0.421 &2.611&0.087&$-2.2$ &  -0.143 &G0 Ib &D  \\      
 HR~7542 &376. &5750.  &2.15   &0.605 & 0.426 &2.633&0.185&$-2.2$ &   0.017 &F8 Ib-II &pAGB?\\  
 HR~7770 &960. &6180.  &1.53   &0.930 & 0.864 &2.665&0.329&$-5.4$ & 0.080& F5 Ib& D \\     
 HR~7796 &562. &5790.  &1.02   &0.818 & 0.425 &2.645&0.026&$-6.6$ &   0.234 &F8 Ib &V,S \\ 
 HR~7823 &1010.&6760.  &1.92   &1.294 & 1.674 &2.729&0.219&$-4.7$ & 0.000& F1 II & \\      
 HR~7834 &235. &6570.  &2.32   &1.022 & 0.259 &2.691&0.060&$-3.1$  & -0.078 &F5 II &V  \\   
 HR~7847 &1040.&6290.  &1.44   &1.022 & 1.044 &2.684&0.295&$-5.9$ & 0.077&F5 Iab & \\  
 HR~8232 &165. &5490.  &1.86   &0.537 & 0.481 &2.600&0.024&$-3.3$ &   0.076 &G0 Ib &D \\ 
 HR~8313 &283. &4910.  &1.58   &0.203 & 0.721 &2.593&$-$0.032&$-2.8$ &  -0.069 &G5 Ib &V \\
 HR~8412 &284. &5280.  &2.35   &0.255 & 0.618 &2.596&0.024&$-1.0$  & -0.066 &G5 Ia &D \\      
 HR~8414 &161. &5210.  &1.76   &0.386 & 0.562 &2.596&0.002&$-3.1$ &   0.020 &G2 Ib &D   \\      
 HR~8692 &413. &4960.  &1.90   &0.210 & 0.662 &2.597&$-$0.008&$-1.8$ &  -0.087 &G4 Ib & \\  
\hline
\end{tabular}
\end{center}
Comments code: V-variable, D-Double, S-$uvby\beta$ from Simbad, pAGB?- possible post-AGB star, RRL- RR Lyrae, SB-Spectroscopic Binary. }
\end{table*}
 
\section {The $T_{\rm \lowercase {eff}}$ calibrations}
\label{temp}

\begin{figure}
\begin{center}
\includegraphics[width=7.5cm,height=7.5cm]{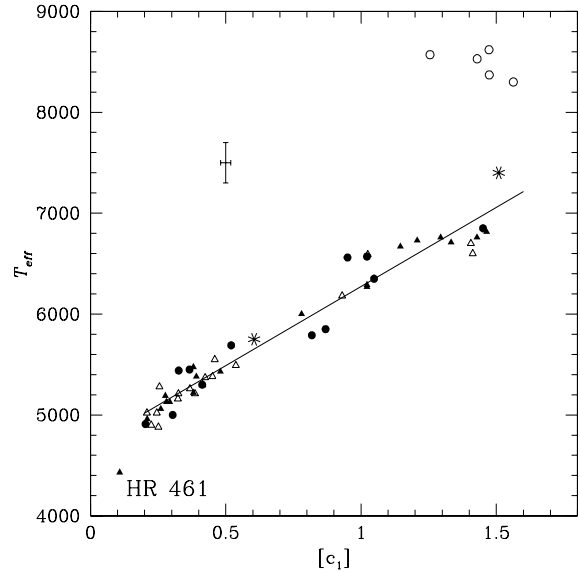}
\caption{Temperature dependence of the reddening free [$c_1$] index. The open circles correspond to five 
A type stars which along with the K1 star HR 461 are not included in the fit. The fit is valid for F-G type stars.
Filled circles represent variable stars, empty triangles double stars and filled triangles non-variables. Asterisks correspond to two possible post-AGB stars in the sample of Lyubimkov et al. (2010).
The error bars correspond to typical uncertainties of $\pm 200$K and $\pm 0.019$ mag in $T_{\rm eff}$
and [$c_1$] respectively.}
    \label{te_bc1}
\end{center}
\end{figure}

\begin{figure}
\begin{center}
\includegraphics[width=7.5cm,height=7.5cm]{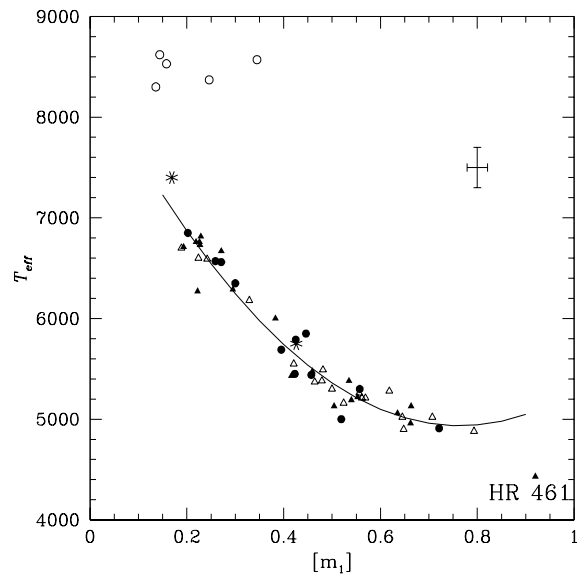}
\caption{Temperature dependence of the reddening free [$m_1$] index. Symbols are as in Fig. \ref{te_bc1}. The error bars correspond to typical uncertainties of $\pm 200$K and $\pm 0.021$ mag in $T_{\rm eff}$ and [$m_1$] respectively.}
    \label{te_bm1}
\end{center}
\end{figure}

Plots of $T_{\rm eff}$ as a function of [$c_1$] and [$m_1$] are shown in Figs. \ref{te_bc1} and 
\ref{te_bm1} respectively. Variable, double and non-variable stars have been plotted with different symbols 
to check if their nature contributes to the scatter or if they appear as outliers.
For the variable stars we used averages from rather scarce photometry not covering the full variational cycles, despite of which they follow the trends as well as the non-variable and double
stars.

The $T_{\rm eff}$-[$c_1$] correlation is good for F-G type stars. Five A-type stars included in the sample of Lyubimkov et al. (2010); HR~825, HR~1740, HR~2874, HR~3183 and  HR~6081 are identified as open circles and the K1 star HR 461 (Yoss 1961)
were not included in the fit. The line has the form:

\begin{equation}
\label{TeBc1} 
T_{\rm eff} = 1566.6 (\pm 50.7) [c_1] + 4704.0 (\pm 40.5),
\end{equation}

\noindent
where the standard deviation is 152~K and the correlation coefficient is $R=0.98$.

We can confirm from the models of Lester, Gray \& Kurucz (1986) that the $c_1$ index is temperature sensitive
for $T_{\rm eff}$ $\leq$ 7000K and that for hotter stars the sensitivity of the index changes and becomes
more gravity dependent. This explains why the A-type stars do not follow the smooth correlation 
displayed by the F-G stars.

The T$_{\rm eff}$-[$m_1$] correlation shows a non-linear form that can be represented by:

$$ T_{\rm eff} = 6081.3 (\pm 817.0) [m_1]^2 - 9294.9 (\pm 731.8) [m_1] $$
\begin{equation}
\label{TeBm1}
+ 8486.7 (\pm 149.5).
\end{equation}

\noindent
Similar to [$c_1$] the five A-type stars and the K1 star HR 461 were not included. The standard deviation is 152~K and the correlation coefficient is $R=0.98$.

In order to explore a possible colour dependence on the above calibrations, we have selected the dereddened colour $(b-y)_0$. To calculate the reddenings 
for the F-G supergiants we used the calibration given in eq. 5 of Arellano Ferro \& Parrao (1990) which provides $E(b-y)$ from ($b-y)$, $c_1$ and $m_1$ indices with an accuracy of about $\pm$0.03 mag. The colour excesses so calculated are given in column 8 of Table \ref{T1}. Since most of these supergiants of low reddening, and given the scatter in 
the reddening calibration, some small negative values are expected. These negative 
values should be used as zero reddening. The resulting equations with the colour term included have 
slightly smaller standard deviations and higher correlation coefficients. They are of the form:

$$T_{\rm eff} = 501.6 (\pm 255.1) [c_1] - 1980.1 (\pm 467.3) (b-y)_0 $$ \begin{equation}
\label{Tec1by}
+ 6252.0 (\pm 367.0),
\end{equation}

\noindent 
with the standard deviation being 129~K and the correlation coefficient $R=0.98$.
\smallskip
Or for the [$m_1$] index:

$$ T_{\rm eff} = 2244.0 (\pm 958.6) [m_1]^2 - 2350.0 (\pm 1407.1) [m_1] $$
\begin{equation}
\label{Tem1by}
-2244.0 (\pm 454.7) (b-y)_0 + 7375.8 (\pm 236.8).
\end{equation}

\noindent 
with the standard deviation being 119~K and the correlation coefficient $R=0.99$.

At this point it is important to remark that since Lyubimkov et al. (2010) used [$c_1$], among other indeces, to estimate $T_{\rm eff}$, the correlations
in Figs. \ref{te_bc1} and \ref{te_bm1} are expected. The calibrations of 
Eqs 1 to 4 are naturally in the temperature scale of Lyubimkov et al. (2010).
What it is offered here are functional relations that can be confortably used
to estimate $T_{\rm eff}$ in F-G supergiants from the $c_1$ photometry with comparable accuracies. 

The H$\beta$ index is also linearly correlated with the temperature. For the linear fit between
H$\beta$ and , the  for the 47 F-G stars with H$\beta$ in Table \ref{T1}, the standard
deviation is 234 K and the correlation coefficient is R=0.93. This
calibration being of less quality than the four represented by eqs. \ref{TeBc1} to \ref{Tem1by}, 
we have not illustrated it. 

\section{The \lowercase {log}~$\lowercase {g}$ calibration}
\label{grav}

Calibrations of the surface gravity in yellow supergiants and bright giants
have been performed in the past by Gray (1991) and Arellano Ferro \& Mendoza (1993). The strategy was to use the [$m_1$]-[$c_1$] plane, on which the locus of a "standard line", determined from F-G Ib supergiants exclusively, was defined 
by Gray (1991). Then, a vertical parameter; $\Delta [c_1] = [c_1] - [c_1](standard~ line)$, can be measured. It has been shown that $\Delta [c_1]$ is correlated with 
log~$g$. Arellano Ferro \& Mendoza (1993) used the spectroscopic gravities for 27 F-G supergiants given 
by Luck \& Bond (1989) to calibrate the correlation. In this work we attempt a new calibration in the light of the new values of log~$g$ given 
by Lyubimkov et al. (2010). The value of $\Delta [c_1]$ for each star in the sample
is given in column 10 of Table \ref{T1}. Fig. \ref{te_Dc1} shows the   
$\Delta [c_1]$ dependence on log~$g$. As before, the five A type and the K1 stars in the sample of Lyubimkov et al. (2010) have not been included. Among the F-G stars three outliers were noted in the 
$\Delta [c_1]$-log~$g$ plane; HR~1242 (F0~II) and HR~3073 (F1~Ia), both stars
are non-variables, and the double star HR~6536. We have no ready explanation for their discordant position and they were ignored. The remaining 47 stars display a clear relationship and the straight line has the form:

\begin{figure}
\begin{center}
\includegraphics[width=7.5cm,height=7.5cm]{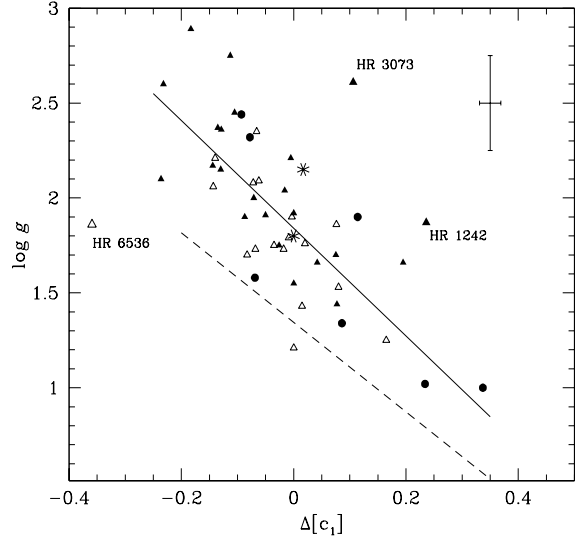}
\caption{Gravity dependence on the index $\Delta [c_1]$. Symbols are as in Fig. \ref{te_bc1}. Segmented line represents the calibration of Arellano Ferro \& Mendoza (1993), which is 0.3-0.5 dex shifted towards lower gravities. See text for discussion. The error bars correspond to typical uncertainties of $\pm 0.25$~dex and $\pm 0.019$~mag in log$~g$
and $\Delta [c_1]$ respectively}
    \label{te_Dc1}
\end{center}
\end{figure}

\begin{figure}
\begin{center}
\includegraphics[width=7.5cm,height=7.5cm]{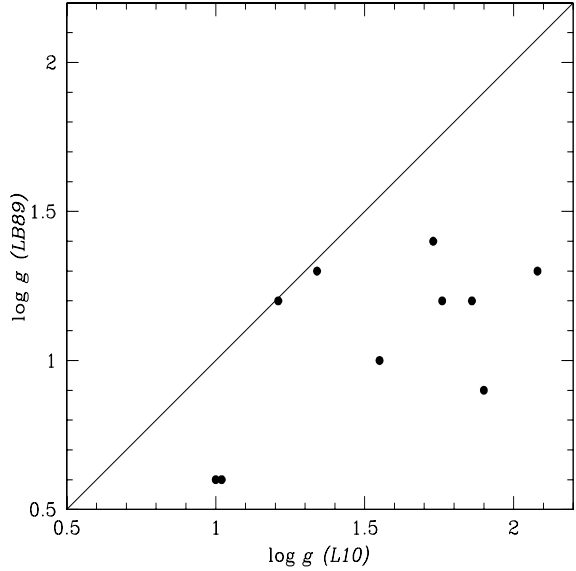}
\caption{Comparison of log~$g$ values for ten stars in common between Lyubimkov et al. (2010) and Luck \& Bond (1989). See text for discussion.}
    \label{LGvsLG}
\end{center}
\end{figure}

\begin{equation}
\label{gc1}
{\rm log}~g = -2.836 (\pm 0.349) \Delta [c_1] + 1.841 (\pm 0.040),
\end{equation}  

\noindent 
with the standard deviation being 0.267 dex and the correlation coefficient is $R=0.77$.

Since there are evidences that log~$g$ depends on $T_{\rm eff}$ for a given luminosity class (Straizys \& Kuriliene 1981) we attempted to incorporate the colour term ($b-y)_0$ but found it non significant.

Eq. \ref{gc1}, when compared with eq. 5 of Arellano Ferro \& Mendoza (1993), it has smaller dispersion and a slightly higher correlation coefficient. However we should note that the present calibration indicates lower gravities by 0.3-0.5 dex for a given $\Delta [c_1]$. This invited a comparison of the gravities used by Arellano Ferro \& Mendoza (1993), taken from Luck \& Bond (1989), with the gravities used in the present work from Lyubimkov et al. (2010) and listed in Table \ref{T1}.
There are ten F and early G type stars in common and in Fig. \ref{LGvsLG} it is shown that the gravities of Luck \& Bond (1989) are systematically smaller. 
Luck \& Bond (1989) found that their spectroscopic gravities, obtained mostly from the 
ionization equilibrium condition on Fe I and Fe II lines, are systematically smaller by 0.3 dex than gravities
derived via the membership of stars in clusters and their photometry. Similar discrepancies were 
found by Luck \& Lambert (1985) for classical cepheids when spectroscopic gravities are compared with 
gravities obtained via PLC relationship. These authors forward a possible explanation in the fact that
models built under the assumption hidrostatic equilibrium may not be a good representation for the extended atmospheres of supergiant stars. The reader is refered to the detailed discussions on this
point given in the above papers, and similar arguments might be invoked to explain the discrepancies
displayed in Fig. \ref{LGvsLG}.

\begin{figure}
\begin{center}
\includegraphics[width=7.5cm,height=7.5cm]{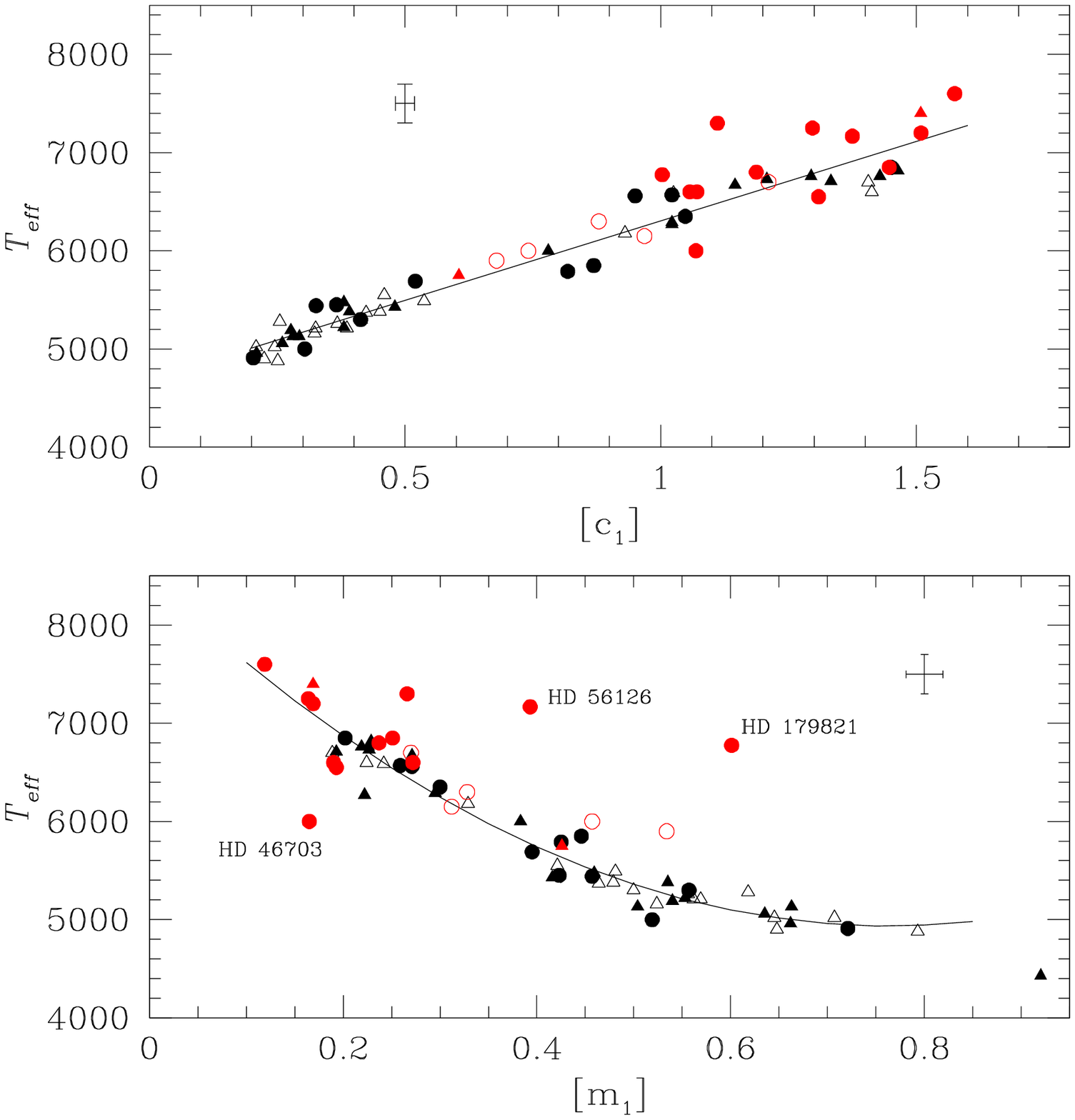}
\caption{Inclusion of postAGB and RV Tau stars in the $T_{\rm eff}$ calibrations. Black symbols are as in Fig. 1. Filled red circles are post-AGB stars and open red circles 
are RV Tau stars. The red triangle is the post AGB candidate star HR~7542 
plotted as an asterisk in Figs. 1, 2 and 3.}

    \label{pAGB}
\end{center}
\end{figure}

\begin{figure}
\begin{center}
\includegraphics[width=7.5cm,height=7.5cm]{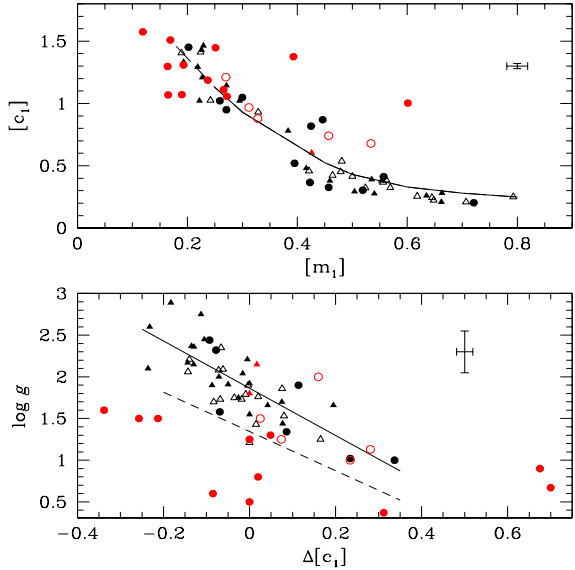}
\caption{Inclusion of postAGB and RV Tau stars in the log~$g$ calibrations. The top panel shows the [$m_1$]-[$c_1$] plane and the fiducial line definded by Gray (1991),
relative to which the $\Delta [c_1]$ parameter is measured. The lower panel shows the 
distribution of the different groups of stars in the $\Delta [c_1]$- log~$g$ plane. Symbols are as in Fig.\ref{pAGB}. See text for discussion.}
    \label{pAGBlg}
\end{center}
\end{figure}

\section{The case of \lowercase {post}-AGB stars}
\label{pagb}

We often have wonder if the calibrations of physical parameters that have been worked out for young
yellow supergiants would be equally valid for more evolved and less massive post-AGB stars
of similar temperature and gravity. We found a few well known post-AGB and RV Tau stars in the catalogue of Stasi\'nska et al. (2008) with reported values of effective temperature and gravity 
estimated by spectroscopic techniques, and with Str\"omgren photometry available in the Simbad database. These stars are listed in Table \ref{T2} and plotted in Fig. \ref{pAGB} 
 with red symbols on the [$c_1$]-$T_{\rm eff}$ and [$m_1$]-$T_{\rm eff}$ planes
 of Figs. \ref{te_bc1} and \ref{te_bm1}. The stars being rather faint, their spectral types may be difficult to determine with high accuracy, for this reason and for the sake 
of not cutting the sample too short, we included late A stars and one K0 star. 
For the [$c_1$] index the post-AGB and RV Tau
stars follow very well the trend defined by the F-G supergiants and therefore it seems that the calibration is also valid for these types of stars stars. We have refrained from yet fitting the
straight line with all points since the difference with eq. 1 would be negligible.
For the [$m_1$] index the post-AGB's and RV Tau also follow the parabola of the F-G supergiants except for three stars HD~46703 (F7~IVw), HD~56126 (F5~Iab:) and 
HD~179821 (G5~Ia). HD~46703 has a spectral type of a subgiant which
may explain its peculiar position, however for HD~56126 and HD~179821 we have no explanation at hand.

\begin{table*}
\footnotesize{
\begin{center}
\caption[{\small }] {\small Physical parameters and photometric indices for a sample of \lowercase{post}AGB and RV Tau stars.}
\label{T2}
\hspace{0.01cm}
 \begin{tabular}{lrccrccl}
\hline
 Star &  $T_{\rm eff}$ & log $g$ &[$c_1$]& [$m_1$]&$\Delta [c_1]$&Sp.T. & Comment \\
 & (K) & (dex)  &  & & & \\
\hline  
HD~46703 &  6000. & 0.40 &  1.069 & 0.165 & -0.463& F7~IVw& pAGB\\      
HD~56126 &  7167. & 0.67 &  1.375 & 0.393 & 0.700& F5~Iab:& pAGB\\      
HD~95767 &  7300. & 1.30 &  1.111 & 0.266 & 0.049& F3~II & pAGB\\    
HD~107369 & 7600. & 1.50 &  1.575 & 0.119 & -0.213& A2~II/III &pAGB\\       
HD~108015 & 6800. & 1.25 &  1.187 & 0.237 & 0.000& F4~Ib/II& pAGB\\      
HD~112374 & 6600. & 0.80  & 1.057 & 0.272 & 0.020& F3~Ia&  pAGB\\         
HD~148743 & 7200. & 0.50  & 1.509 & 0.169 & 0.000& A7~Ib  & pAGB\\    
HD~161796 & 6850. & 0.37 &  1.447 & 0.251 & 0.312&F3~Ib&  pAGB\\
HD~163506 & 6550. & 0.60 &  1.309 & 0.193 & -0.085& F2~Ibe& pAGB\\      
HD~172481 & 7250. & 1.50  & 1.297 & 0.164 & -0.257& F2~Ia0 & pAGB\\
HD~179821 & 6775.&  0.90 &  1.003 & 0.601 & 0.675& G5~Ia & pAGB\\        
HD~190390 & 6600. & 1.60 &  1.071 & 0.190 & -0.338& F1~III&pAGB\\
\hline    
HD~170756 & 5900. & 1.13  & 0.679 & 0.534 & 0.281& K0~III& RV Tau\\     
AR~Pup  &   6300. & 1.50  & 0.879 & 0.328 & 0.025& F0~Iab&    RV Tau\\   
IW~Car  &   6700. & 2.00  & 1.211 & 0.270 & 0.160& A4~Ib/II& RV Tau\\      
RU~Cen  &   6000. & 1.00  & 0.741 & 0.457 & 0.234& F7/F8;A4~Ib & RV Tau\\     
EN~TrA  &   6150. & 1.25  & 0.968 & 0.312 & 0.074& F2~Ib&RV Tau\\
\hline
\end{tabular}
\end{center}
}
\end{table*}

Let us now explore, in a similar way, the role of post-AGB stars on the 
$\Delta [c_1]$- log~$g$ plane. As for the of the young supergiants we started by plotting 
the stars on the [$m_1$]-[$c_1$] from where one can estimate $\Delta [c_1]$. Fig. \ref{pAGBlg} 
illustrates both these planes. Except for the already noted howler stars 
HD~46703, HD~56126 and HD~179821, both the postAGB and RV Tau stars follow the 
distribution on the [$m_1$]-[$c_1$] plane. However on the $\Delta [c_1]$- log~$g$ plane
the post-AGB stars are at odds with the distribution of the supergiant stars.
The five RV Tau stars (empty circles) seem to follow the supergiant trend rather well. With the available 
information we cannot say whether the post-AGB stars follow a completely different relationship 
or if this is a result of
the gravities reported by Stasi\'nska et al. (2008) being anomalously too small. In either
case the relationship in eq. 5 is to be used only for the young supergiant stars.

\section{Comments on the determination of $M_V$}
\label{MV}

It would be desirable to posses a calibration that can predict the luminosity of 
young yellow supergiants given an observational parameter, such as for example a
colour index. Since these stars are luminous they could be used as distance indicators. We have attempted before to calculate such calibration 
(Arellano Ferro \& Parrao 1991; Arellano Ferro \& Mendoza 1993), however an accurate result has been rather elusive. 
On the other hand a neat calibration 
of $M_V$ from the intensity of the OI 7774 triplet, valid for a vast range of luminosities, 
was calculated by Arellano Ferro et al. (2003) wich predicts $M_V$ with an uncertainty of 0.38 mag.

\begin{figure}
\begin{center}
\includegraphics[width=7.5cm,height=7.5cm]{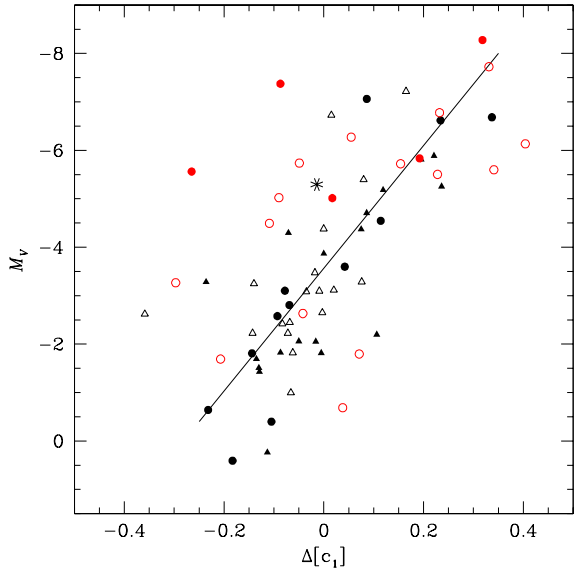}
\caption{Trend of $M_V$ with $\Delta [c_1]$. Black symbols are as in Fig. \ref{te_bc1}. The red symbols correspond to a sample of F-G supergiants (open circles) and well documented postAGB stars (filled circles) taken from Arellano Ferro et al's. (2003) Table 3, and whose values of $M_V$ were calculated from the OI~7774 feature.}
    \label{MVDc1}
\end{center}
\end{figure}

The distances provided by Lyubimkov et al. (2010) could in principle serve to explore 
a calibration of $M_V$ in terms of a photometric index. The absolute magnitudes listed in colum 9 of Table \ref{T1} were calculated from the distances (column 2) and the reddenings (column 8).
Since $\Delta [c_1]$ is correlated with log~$g$ one would expect a relationship between $M_V$ and $\Delta [c_1]$.
In Fig. \ref {MVDc1}  the plot of $\Delta [c_1]$ vs. $M_V$ is shown. Although a trend is clearly seen, the scatter is large ($\sigma$ = 1.3 mag). Inclusion of colour term, $(b-y)_o$ does not improve the relationship  ($\sigma$ = 1.1 mag).

We find of interest to considere a sample of F-G supergiants and 5 well
documented post-AGB stars listed by
Arellano Ferro et al. (2003) (their Table 3) for which their values of $M_V$
have been calculated from the OI~7774 calibration. $uvby\beta$ photometry  for
all these stars is available in the Simbad data base and then we could calculate
their $\Delta [c_1]$ parameter and plot them on Fig. \ref {MVDc1} (red symbols) 
along with the F-G supergiants in Table \ref{T1}. We see that both groups of stars
follow the same trend and display a similar scatter. One can conclude from this  that while 
the distances in Lyubimkov et al. (2010) are consistent with the absolute magnitudes
predicted from the strengths of the OI~7774 feature in F-G supergiants, 
a reliable calibration of $M_V$ in terms of a photometric index, for example $\Delta [c_1]$, is not foreseen.

\section{Conclusions}
\label{concl}

New accurate values of $T_{\rm eff}$ and log~$g$ (Lyubimkov et al. 2010)
allow reliable functional relationships from Str\"omgren reddening free
indices [$c1$] and [$m1$] that allow to predict the effective temperature
and gravity in F-G supergiants. The temperature calibrations are given in eqs. \ref{TeBc1} and \ref{TeBm1} which predict the effective temperature with an 
standard deviation of 152~K. If the reddening is available, alternative calibrations, given by eqs. \ref{Tec1by}  and \ref{Tem1by} provide sligtly smaller standard deviations of 129 and 119~K respectively. These calibrations are valid for the more evolved post-AGB of similar temperature and gravity. Given the fact that RV Tau stars are variables of large amplitude, it is rewarding
to see that the five RV Tau stars included follow the temperature trends very closely. Perhaps their mean temperature and
photometric indices are not too different from the temperatures reported by Stasi\'nska et al. (2008) and the mean photometric indices calculated here.

The stellar gravity can also be predicted from the  $\Delta [c_1]$ index (Gray 1991)
from the calibration of eq. \ref{gc1} with an accuracy 0.28 dex. The colour term was found to be non-significant. This equation is valid for F-G yellow supergiants and very likely for RV Tau stars. However well known post-AGB stars of similar temperature and gravities (see Table 2) do not follow the calibration.

We have not been able to determine a reliable calibration that can predict the absolute magnitude $M_V$ in F-G supergiant stars from photometric indices as
accurately as from the strength of the OI 7774 feature.

\acknowledgments
Numerous comments and suggestions made by Prof. Sunetra Giridhar and of an anonymous referee are thankfully acknowledged.
 This work was supported by DGAPA-UNAM grant through project IN114309. This research has made extensive use of the Simbad
database operated at CDS, Strasbourg, France and of the NASA ADS Astronomy
Query Form.

\end{document}